# A general theory of intertemporal decision-making and the perception of time

**Vijay Mohan K Namboodiri[1], Stefan Mihalas[2], Tanya Marton[1], Marshall G Hussain Shuler[1]***

[1] Dept of Neuroscience, Johns Hopkins University, 725 N Wolfe Street, 914 WBSB, Baltimore, MD 21205, USA

[2] Allen Institute for Brain Science, 551 N 34th St #200, Seattle, WA 98103, USA

* Corresponding Author: shuler@jhmi.edu, +1-410-502-1612

**Animals and humans make decisions based on their expected outcomes. Since relevant outcomes are often delayed, perceiving delays and choosing between earlier versus later rewards (intertemporal decision-making) is an essential component of animal behavior. The myriad observations made in experiments studying intertemporal decision-making and time perception have not yet been rationalized within a single theory. Here we present a theory—Training-Integrated Maximized Estimation of Reinforcement Rate (TIMERR)—that explains a wide variety of behavioral observations made in intertemporal decision-making and the perception of time. Our theory postulates that animals make intertemporal choices to optimize expected reward rates over a limited temporal window; this window includes a past integration interval (over which experienced reward rate is estimated) and the expected delay to future reward. Using this theory, we derive a mathematical expression for the subjective representation of time. A unique contribution of our work is in finding that the past integration interval directly determines the steepness of temporal discounting and the nonlinearity of time perception. In so doing, our theory provides a single framework to understand both intertemporal decision-making and time perception**.

### Introduction

Survival and reproductive success depends on beneficial decision-making. Such decisions are guided by judgments regarding outcomes which are represented as expected reinforcement amounts. As actual reinforcements are often available only after a delay, measuring delays and attributing values to reinforcements that incorporate the cost of time is an essential component of animal behavior[1,2]. Yet, how animals perceive time and assess the worth of delayed outcomes— the quintessence of intertemporal decision-making—though fundamental, remains to be satisfactorily answered[1,3,4]. Rationalizing both the perception of time and the valuation of outcomes delayed in time in a unified framework would significantly improve our understanding of basic animal behavior, with wide-ranging applications in fields such as economics, ecology, psychology, cognitive disease and neuroscience.

In the past, many theories including Optimal Foraging Theory[1,2] (OFT), Discounted Utility Theory[3–5] (DUT), Ecological Rationality Theory[1,6,7] (ERT), as well as other psychological models[3,4,8,9] have been proposed as solutions to this question. Of these, OFT, DUT and ERT attempt to understand ultimate causes of behavior through general optimization criteria, whereas psychological models attempt to understand its proximate biological implementation. The algorithms specified by these prior theories and models for intertemporal decision-making are all defined by their temporal discounting function—the ratio of subjective value of a delayed reward to the subjective value of the reward when presented immediately. These algorithms come in two major forms: hyperbolic (and hyperbolic-like) discounting functions (e.g. OFT and ERT)[1–4], and exponential (and exponential-like, e.g. $\beta$-$\delta$[4,8,9]) discounting functions (e.g. DUT)[3–5]. Hyperbolic discounting functions have been widely considered to be better fits to behavioral data than exponential functions[3,4].

The best-fit parameters of these algorithms to behavioral data that measure discounting steepness, however, lack a biological meaning, and observed changes in their values are left unexplained. For instance, there are many "anomalies" (so-termed in economics) observed in discounting steepness, including, but not limited to, "Magnitude Effect"[3,4,10], "Sign Effect"[3,4] and differential treatment of losses[3,4], that have not yet been explained by a general theory[1–4,8,9,11]. Further, while intertemporal decision-making necessarily requires perception of time, theories of intertemporal decision-making and time perception[12,13] are largely independent and do not attempt to rationalize both within a single framework.

We argue that the inability of prior theories to rationalize behavior stems from the lack of biologically-realistic constraints on general optimization criteria (see *Supplementary Information-1.1*). While psychological models[3,4,8,9] are biologically-grounded by design, they lack the parsimony of ultimate theories. In contrast, existing ultimate theories[1–7] lack the biological plausibility of proximate psychological models (*Supplementary Figure S1*). The motivation for our present work was to create a biologically-realistic and parsimonious theory of intertemporal decision-making and time perception which proposes an algorithmically-simple decision-making process to maximize fitness. Our TIMERR theory proposes an algorithm for intertemporal choice that aims to maximize expected reward rate based on, and constrained by, memory of past reinforcement experience. As a consequence, it postulates that time is subjectively represented such that representations of changes in expected reward rate are accurate. In doing so, we are capable of explaining a wide variety of fundamental observations made in intertemporal decision-making and time perception. These include hyperbolic discounting[2–4,6], "Magnitude"[3,4,10] and "Sign" effects[3,4], differential treatment of losses[3,4], scaling of timing errors with interval duration[12–15], and, observations that impulsive subjects (as defined by abnormally steep discounting) under-produce[16] time intervals and show larger timing errors[16,17] (for a full list, see *Supplementary Information-2.4.1*). It thereby recasts the above-mentioned "anomalies" not as flaws, but as features of reward-rate optimization under experiential constraints.

## Results

To illustrate the motivation and reasoning behind our theory, we consider a simple behavioral task. In this task, an animal must make decisions on every trial between two randomly chosen (among a finite number of possible alternatives) reinforcement-options. An example environment with three possible reinforcement-options is shown in Figure 1a. We assert that the goal of the animal is to gather the maximum total reward over a fixed amount of time, equivalently, to attain the maximum total (global) reward rate over a fixed number of trials.

Assuming a stationary reinforcement-environment in which it is not possible to directly know the pattern of future reinforcements, an animal may yet use its past reinforcement experience to instruct its current choice. Provisionally, suppose also that an animal can store its entire reinforcement-history in the task in its memory. It can then maximize the total reward rate achieved by the end of a current trial by picking the option that when chosen, would lead to the *highest global reward rate over all trials until, and including, the current trial*, i.e.

$$\text{Pick option with the highest value for } \frac{R+r_i}{T+t_i} \quad (\textit{Equation 1})$$

where $T$ is the total time elapsed in the session so far, $R$ is the total reward accumulated so far and ($r_i$, $t_i$) is the reward magnitude and delay, respectively, for the various reinforcement-options on the current trial.

Under the above conditions, this algorithm yields the highest possible reward rate achievable at the end of any given number of trials. In contrast, previous algorithms for intertemporal decision-making, while being successful at fitting behavioral data, fail to maximize global reward rates. For the example reinforcement-environment shown in Figure 1a, simulations show that the algorithm in *Equation 1* outperforms other extant algorithms (hyperbolic discounting, exponential discounting, two-parameter discounting) by more than an order of magnitude (Figure 1b).

The reason why extant alternatives fare poorly is that they do not account for opportunity cost, i.e. the cost incurred in the lost opportunity to obtain better rewards than currently available. In the example considered, two of the reinforcement-options are significantly worse than the third (Figure 1c). Hence, in a choice between these two options, it is even worth incurring a small punishment ($-0.01) at a short delay for sooner opportunities of obtaining the best reward ($5) (Figure 1c). Previous models, however, pick the reward ($0.1) in favor of the punishment since they do not have an estimate of opportunity cost. In contrast, by storing the reinforcement history, *Equation 1* accounts for the opportunity cost, and picks the punishment. Recent experimental evidence suggests that humans indeed accept small temporary costs in order to increase the opportunity for obtaining larger gains[18].

The behavioral task shown in Figure 1a is similar to standard laboratory tasks studying intertemporal decisions[1,3,4,19]. However, in naturalistic settings, animals commonly have the ability to forgo any presented option. Further, the number of options presented on a given trial can vary and could arise from a large pool of possible options. An illustration of such a task is displayed in Figure 1d, showing the outcomes of five past decisions. Decision 2 illustrates an instance of incurring an opportunity cost. Decision 3 shows the presentation of a single option that was forgone, leading to the presentation of a better option in decision 4. Though the options presented in decision 5 are those in decision 1, the animal's choice behavior is the opposite, as a result of changing estimations of opportunity cost. Results of performance in such a simulated task (with no punishments) are shown in Figure 1e, again showing *Equation 1* outperforming other models (see *Methods*).

It is important to note that while the extent to which *Equation 1* outperforms other models depends on the reinforcement-environment under consideration, its performance in a stationary environment will be greater than or equal to previous decision models. However, biological systems face at least three major constraints that limit the appropriateness of *Equation 1*: 1) their reinforcement-environments are non-stationary; 2) integrating reinforcement-history over arbitrarily long intervals is computationally implausible, and, 3) indefinitely long intervals without reward cannot be sustained by an animal (while maintaining fitness) even if they were to return the highest long-term reward rate (e.g. choice between 100000 units of food in 100 days vs. 10 units of food in 0.1 day). Hence, in order to be biologically-realistic, TIMERR theory states that the interval over which reinforcement-history is evaluated, the past-integration-interval ($T_{ime}$ ; *ime* stands for *in my experience*), is finite. Thus the TIMERR algorithm states that animals maximize reward rates over an interval including $T_{ime}$ and the learned expected delay to reward (*t*) (see TIMERR algorithm in Figure 2a-b; *Supplementary Information-1.1*). This modification renders the decision algorithm shown in *Equation 1* biologically-plausible.

Therefore, the TIMERR algorithm acts as a temporally-constrained, experience-based, solution to the optimization problem of maximizing reward rate. It requires that only experienced magnitudes and times of the rewards following conditioned stimuli are stored, therefore predicting that intertemporal decisions of animals will not incorporate post-reward delays due to limitations in associative learning[6,11,20,21]. consistent with prior experimental evidence showing the insensitivity of choice behavior to post-reward delays[1,3,6,11,21] (see *Supplementary Information-2.1.2* for a detailed discussion). Indirect effects of post-reward delays on behavior[21] can however be explained as resulting from the implicit effect of post-reward delays on past reward rate.

From the TIMERR algorithm, it is possible to calculate the subjective value of a delayed reward (Figure 2c)—defined as the amount of immediate reward that is subjectively equivalent to the delayed reward (*Supplementary Information-1.2*). This expression is given by

$$SV(r,t) = \frac{r - a_{est} t}{1 + \frac{t}{T_{ime}}}$$  (*Equation 2*)

where $a_{est}$ is an estimate of the average reward rate in the past over the integration window $T_{ime}$ with the reward option specified by a magnitude *r* and a delay *t*.

*Equation 2* presents an alternative interpretation of the algorithm: the animal is estimating the net worth of pursuing each delayed reward by subtracting the opportunity cost incurred by forfeiting potential alternative reward options during the delay to a given reward. This is because the numerator in *Equation 2* represents the expected reward gain but subtracts this opportunity cost, $a_{est} t$, which corresponds to a baseline expected amount of reward that might be acquired over *t*. The denominator is the explicit temporal cost of waiting (*Supplementary Information-1.2*).

The temporal discounting function—the ratio of subjective value to the subjective value of the reward when presented immediately—is given by (based on *Equation 2*)

$$D(r,t) = \frac{SV(r,t)}{r} = \frac{1 - \frac{a_{est}}{r}t}{1 + \frac{t}{T_{ime}}} \quad (Equation\ 3)$$

This discounting function is hyperbolic with an additional, dynamical (changing with $a_{est}$) subtractive term. The effects of varying the parameters, viz. the past integration interval ($T_{ime}$), estimated average reward rate ($a_{est}$) and reward magnitude (*r*), on the discounting function are shown in Figure 3. The steepness of this discounting function is directly governed by $T_{ime}$, the past integration interval (Figure 3a). In other words, the longer one integrates over the past to estimate reinforcement history, the higher the tolerance to delays when considering future rewards, thus rationalizing abnormally steep discounting (characteristic of impulsivity) as resulting from abnormally low values of $T_{ime}$. As opportunity costs ($a_{est}$) increase, delayed rewards are discounted more steeply (Figure 3b). Also, as the magnitude of the reward increases (Figure 3c), the steepness of discounting becomes lower, referred to as the "Magnitude Effect"[3,4,10] in prior experiments. Further, it is shown that gains are discounted more steeply than losses of equal magnitudes in net positive environments (Figure 3d), as shown previously and referred to as the "Sign Effect"[3,4]. It must also be pointed out that the discounting function for a loss becomes steeper as the magnitude of the loss increases, observed previously as the reversal of the "Magnitude Effect" for losses[22]. In fact, when forced to pick a punishment in a net positive environment, low-magnitude (below $a_{est} \times T_{ime}$) losses will be preferred immediately while higher-magnitude losses will be preferred when delayed (Supplementary Figure S3), as has been experimentally observed[3,4,22] (for a full treatment of the effects of changes in variables, see *Supplementary Information-2.1.1*).

Attributing values to rewards delayed in time necessitates representations of those temporal delays. These representations of time are subjective, as it is known that time perception varies within and across individuals[12–16], and that errors in representation of time increase with the interval being represented[12–15]. Since TIMERR theory states that animals seek to maximize expected reward rates, we posit that time is represented subjectively (Figure 4a) so as to result in accurate representations of *changes* in expected reward rate. In other words, subjective time is represented so that subjective reward rate (subjective value/subjective time) equals the true expected reward rate less the baseline expected reward rate ($a_{est}$). Hence, if the subjective representation of time associated with a delay *t* is denoted by *ST(t)*,

$$\frac{SV(r,t)}{ST(t)} = \left(\frac{r}{t} - a_{est}\right) \quad (Equation\ 4)$$

Combining *Equation 4* with *Equation 2*, we get

$$ST(t) = \frac{t}{1 + \dfrac{t}{T_{ime}}} \quad\quad (Equation\ 5)$$

Such a representation has the property of being bounded ($ST(\infty) = T_{ime}$), thereby making it possible to represent very long durations within the finite dynamic ranges of neuronal firing rates. Plots of the subjective time representation of delays between one and sixty seconds are shown in Figure 4b for two different values of $T_{ime}$. As mentioned previously (Figure 3a), a lower value of $T_{ime}$ corresponds to steeper discounting, characteristic of more impulsive decision-making. It can be seen that the difference in subjective time representations between 40 and 50 seconds is smaller for a lower $T_{ime}$ (high impulsivity). Hence, higher impulsivity corresponds to a reduction in the ability to discriminate between long intervals (Figure 4a-b).

Internal time representation has been previously modeled using accumulator models[15] that incorporate the underlying noisiness in information processing. Using a simple noisy accumulator model (see *Methods,* Figure S2) that represents subjective time according to *Equation 5*, we simulated a time interval reproduction task[12,15], the results of which are shown in Figure 4c-d. Lower values of $T_{ime}$ correspond to an underproduction of time intervals, with the magnitude of underproduction increasing with increasing durations of the sample interval (Figure 4c). When attempting to reproduce a 60 s sample interval, the magnitude of underproduction decreases with increases in $T_{ime}$, or equivalently, with decreasing impulsivity (Figure 4d). These predictions are supported by prior experimental evidence[16].

Prior studies have observed that the error in representation of intervals increases with their durations[12–15]. Such an observation is consistent with the subjective time representation presented here (Figure 4a-b). TIMERR theory predicts that the representation errors will be larger when $T_{ime}$ is smaller (higher impulsivity) (Figure 4a-b), as observed experimentally [16,17]. Prior studies investigating the relationship between time duration and reproduction error have observed a linear scaling ("scalar timing") within a limited range [12–15]. Though the simple accumulator model considered here predicts a quadratic scaling, it nonetheless appears linear within a limited range (*Supplementary Information-1.5*). Moreover, the specific prediction of a U-shaped coefficient of variation (spread/central tendency) for the production times within a single subject (*Equation S.12*) accords with experimental evidence examining a wider range of sample durations[13,23].

Time perception is also studied using temporal bisection experiments[12,24,25] in which subjects categorize a sample interval as closer to a short or a long reference interval. *Equation 5* predicts that the sample duration at which there is maximum uncertainty—the bisection point— will be between the arithmetic and harmonic means of the reference durations (*Supplementary Information-1.4*), depending on the value of $T_{ime}$ (*Equation S.11*). Experimental evidence is consistent with this prediction[24–26] as well as with the related prediction that the steeper the temporal discounting, the lower the bisection point[24].

**Discussion**

Our theory provides a simple algorithm for decision-making in time. The algorithm of TIMERR theory, in its computational simplicity, could explain results on intertemporal choice observed across the animal kingdom[2–4], from insects to humans. Higher animals, of course, could evaluate subjective values with greater sophistication to build better models of the world including predictable statistical patterns of the environment and estimates of risks involved in waiting (*Supplementary Information-1.3*). It must also be noted that other known variables influencing subjective value like satiety[2,27], the non-linear utility of reward magnitudes[2,27] and the non-linear dependence of health/fitness on reward rates[2] have been ignored. Such factors, however, can be included as part of an extension of TIMERR theory while maintaining its inherent computational simplicity. We derived a generalized expression of subjective value that includes such additional factors (*Supplementary Equation S.10*), capturing even more variability in observed experimental results[3,4] (*Supplementary Information-1.3.3-1.3.6*). It must also be noted that while we have ignored the effects of variability in either delays or magnitudes, explanations of such effects have previously been proposed[20,28] and are not in conflict with our theory.

In environments with time-dependent changes of reinforcement statistics, animals should have an appropriately sized past integration interval depending on the environment so as to appropriately estimate opportunity costs (e.g. integrating reward-history from the onset of winter would be highly maladaptive in order to evaluate the opportunity cost associated with a delay of an hour in the summer; also see *Supplementary Information-2.2*). In keeping with the expectation that animals can adapt past integration intervals to their environment, it has been shown that humans can adaptively assign different weights to previous decision outcomes based on the environment[29,30]. As *Equations 2 and 3* show (Figure 3a), changes in $T_{ime}$ would correspondingly affect the steepness of discounting. This novel prediction has two major implications for behavior: 1) *the discounting steepness of an individual need not be a constant*, as has sometimes been implied in prior literature[4]; 2) *the longer the past integration interval, the higher the tolerance to delays when considering future rewards*. In accordance with the former prediction, several recent reviews have suggested that discounting rates are variable within and across individuals[4,9,19,31,32]. The latter prediction states that impulsivity[33], as characterized by abnormally steep discounting, could be the result of abnormally short windows of past reward rate integration. This may explain the observation that discounting becomes less steep as individuals develop in age[8], should the longevity of memories increase over development. Further, *Equation 5* states that changes in $T_{ime}$ would lead to corresponding changes in subjective representations of time. Hence, we predict that perceived durations may be linked to experienced reward environments, i.e. "time flies when you're having fun".

Reward magnitudes and delays have been shown to be represented by neuromodulatory and cortical systems[34–36], while neurons integrating cost and benefit to represent subjective values have also been observed[37,38]. Recent reward rate estimation ($a_{est}$) has been proposed to be embodied by dopamine levels over long time-scales[39]. Interestingly, it has been shown that administration of dopaminergic agonists (antagonists) leads to underproduction (overproduction)[40] of time intervals, consistent with a relationship between recent reward rate estimation and subjective time representation as proposed here. Average values of foraging environment have also been shown to be represented in the anterior cingulate cortex[18]. In light of these experimental observations neurobiological models have previously proposed that decisions, similar to our theory, result from the net balance between values of the options currently under consideration and the environment as a whole[18,38]. However, these models do not propose that the effective interval ($T_{ime}$) over which average reward rates are calculated directly determines the steepness of temporal discounting.

TIMERR theory is the first unified theory of intertemporal choice and time perception to capture such a wide array of experimental observations including, but not limited to, hyperbolic discounting[2–4,6], "Magnitude" [3,4,10] and "Sign" effects[3,4], differential treatment of losses[3,4], as well as correlations between temporal discounting, time perception[16], and timing errors[12–17] (see *Supplementary Information-2.4.1* for a full list). While the notion of opportunity cost long precedes TIMERR, TIMERR's unique contribution is in stating that the past integration interval over which opportunity cost is estimated directly determines the steepness of temporal discounting and the non-linearity of time perception. As a direct result, TIMERR theory suggests that the spectra of aberrant timing behavior seen in cognitive/behavioral disorders[15–17] (Parkinson's disease, schizophrenia and stimulant addiction) can be rationalized as a consequence of aberrant integration over experienced reward history. Hence, TIMERR theory has major implications for the study (*Supplementary Information-2.1-2.4*) of decision-making in time and time perception in normal and clinical populations.

**Methods**

All simulations were run using MATLAB R2010a.

*Simulations for Figure 1*:

Figure 1b: Each of the four decision-making agents ran a total of hundred trials. This was repeated ten times to get the mean and standard deviation. Every trial consisted of the presentation of two reinforcement-options randomly chosen from the three possible alternatives as shown in Figure 1a.

Figure 1e: The following four possible reward-options were considered, expressed as (*r,t*): (0.1,100), (0.0001,2), (5,2), (5,150). The units are arbitrary. To create the reinforcement-environment, a Poisson-process was generated for the availability-times of each of the four options. These times were binned into bins of size 1 unit, such that each time bin could consist of zero to four reward-options. The rate of occurrence for each option was set equally to 0.2 events/unit of time. For the three previous decision-making models, the parameters were tuned for maximum performance by trial and error. Forgoing an available reward-option was not possible for these models since their subjective values are always greater than zero for rewards.

*Simulations for Figure 4:*

An accumulator model described by the following equation was used for simulations of a time reproduction task.

$$dST(t) = \frac{dt}{\left(1+\frac{t}{T_{ime}}\right)^2} + \sigma\, dW_t$$

where $W_t$ is a standard Wiener process and $\sigma$ is the magnitude of the noise. $\sigma$ was set to 10%. This equation is consistent with the subjective time expression shown in *Equation 5*.

Every trial in the time reproduction task consisted of two phases: a time measurement phase and a time production phase. During the time measurement phase, the accumulator integrates subjective time until the expiration of the sample duration (Figure S2). The subjective time value at the end of the sample duration is stored in memory as the threshold for time production, i.e.

$$Threshold(t) = ST(t)$$

During the time production phase, the accumulator integrates subjective time until the threshold is crossed for the first time. This moment of first crossing represents the action response indicating the end of the sample duration, i.e.

$$Reproduced\ interval = t : ST(t) \geq Threshold(t)$$

Sample interval durations ranged between 1 and 60 seconds over bins of one second. A total of 2000 trials were performed for each sample duration and $T_{ime}$ to calculate the median production interval as shown in Figure 4c-d. The time bin for integration *dt* was set to be 1/1000 of the sample duration and the integration was carried out up to a maximum time equaling ten times the sample duration.

**Acknowledgements**

We thank Dr. Peter Holland, Dr. Veit Stuphorn, Dr. James Knierim, Dr. Emma Roach, Dr. Camila Zold, Josh Levy, Gerald Sun, Grant Gillary, Jeffrey Mayse, Naween Anand, Arnab Kundu and Kyle Severson for discussions and comments on the manuscript. This work was funded by NIMH (R01 MH084911 and R01 MH093665) to M.G.H.S.

V.M.K.N, S.M and M.G.H.S conceived of the study. V.M.K.N and S.M developed TIMERR theory and its extensions. V.M.K.N ran the simulations comparing the performance of Equation 1 with other models shown in Figure 1b&e and simulations of the time interval reproduction task used to generate Figure 4. T.M and M.G.H.S were involved in intellectual discussions throughout the work. V.M.K.N wrote the manuscript with assistance from S.M, T.M and M.G.H.S. The authors declare no competing financial interests.

**Figures:**

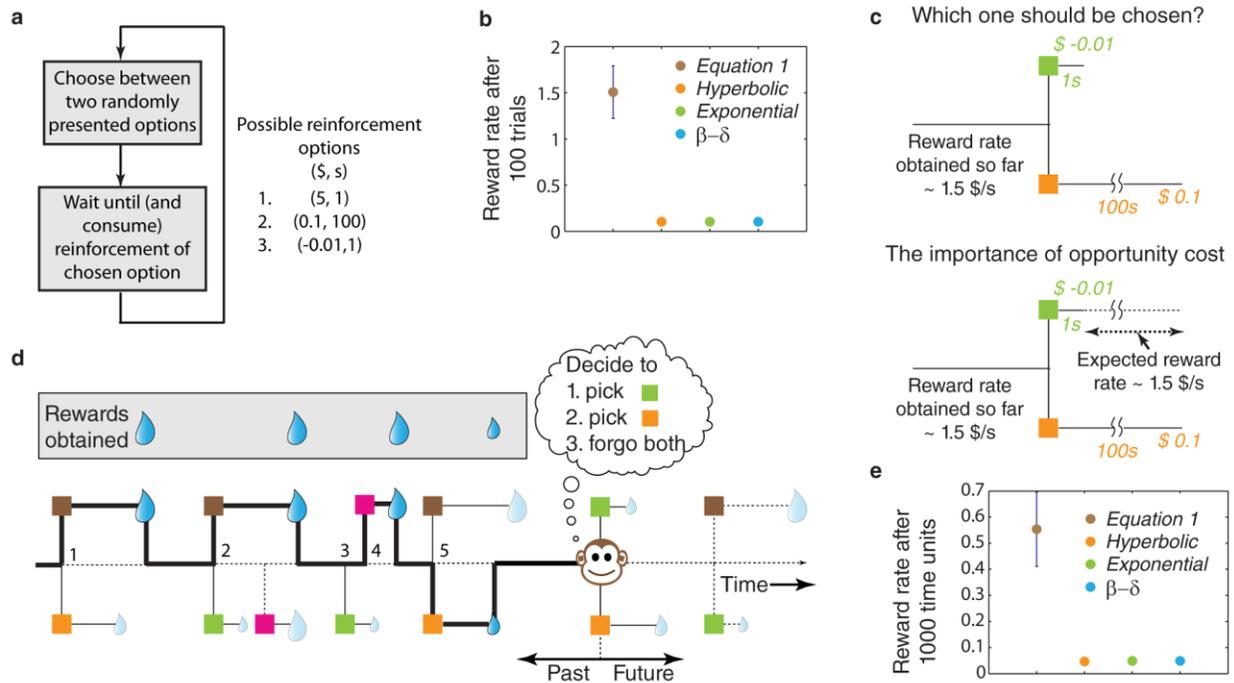

**Figure 1:** *A schematic illustrating the problem of intertemporal decision-making and the rationale for our solution:* **a)** Flow chart of a simple behavioral task, showing the possible reinforcement options. **b)** The performance of four decision-making agents using the four decision processes as shown in the legend (see *Methods*). The parameters of the three previous models were tuned to attain maximum performance. The error bar shows standard deviation. Since the decision rules of these models operate only on the current trial, the corresponding performances have no variability and hence, their standard deviations are zero. **c)** Illustration of the reason for performance failure, showing a choice between the two worst options. The reward rate so far is much higher than the reward rates provided by the two options under consideration. Since these models do not include a metric of opportunity cost, they pick ($0.1, 100s). However, on an average, choosing ($-0.01,1s) will provide a larger reward at the end of 100s. **d)** A schematic illustrating a more natural behavioral task, with choices involving one or two options chosen from a total of four known reinforcement-options. The choices made by the animal are indicated by the bold line and are numbered 1-5. Here, we assume that during the wait to a chosen reinforcement-option, other reinforcement-options are not available (see *Supplementary Information-1.3.4* for an extension). Reinforcement-options connected by dotted lines are unknown to the animal either because they are in the future, or because of the choices made by the animal in the past. For instance, deciding to pursue the brown option in the second choice causes the animal to lose a large reward, the presence of which was unknown at the moment of decision. **e)** Performance of the models in an example environment as shown in **d** (see *Methods* for details). Error bars for the previous models are not visible at this scale. For the environment chosen here, a hyperbolic model (mean reward rate = 0.0465) is slightly worse than exponential and β-δ models (mean reward rate = 0.0490).

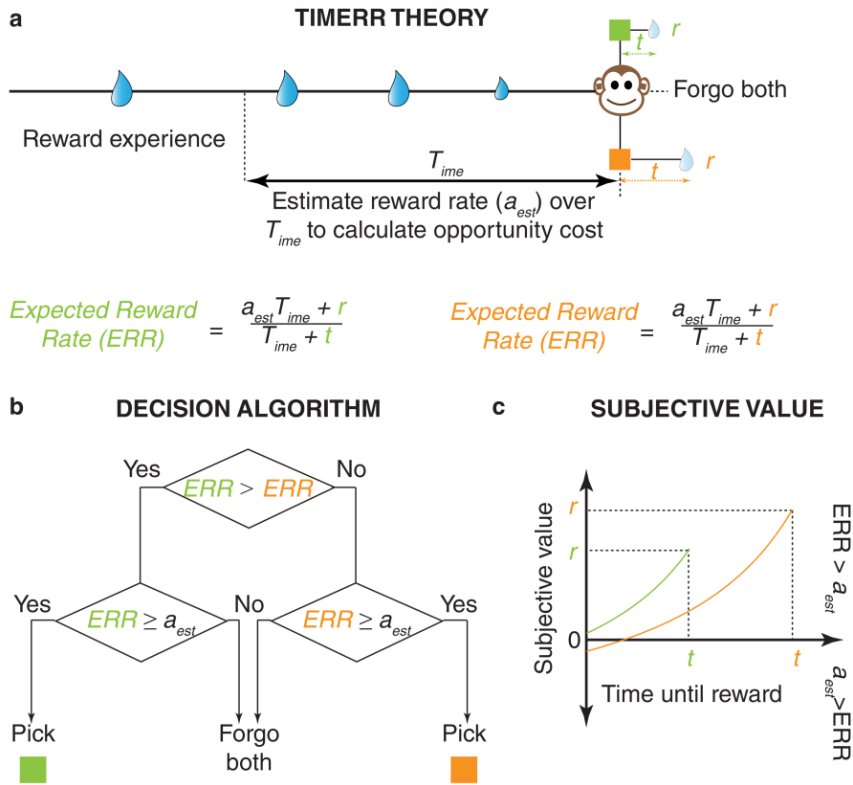

**Figure 2:** *Solution to the problem of intertemporal choice as proposed by TIMERR theory.* **a**) Past reward rate is estimated ($a_{est}$) by the animal over a time-scale of $T_{ime}$ (*Supplementary Information-2.3*). This estimate is used to evaluate whether the expected reward rates upon picking either current option is worth the opportunity cost of waiting. **b**) The decision algorithm of TIMERR theory shows that the option with the highest expected reward rate is picked, so long as this reward rate is higher than the past reward rate estimate ($a_{est}$). Such an algorithm automatically includes the opportunity cost of waiting in the decision. **c**) The subjective values for the two reward options shown in **a** (time-axis scaled for illustration) as derived from the decision algorithm (*Equation 2*) are plotted. In this illustration, the animal picks the green option. It should be noted that even if the orange option were to be presented alone, the animal would forgo this option since its subjective value is less than zero. Zero subjective value corresponds to $ERR = a_{est}$.

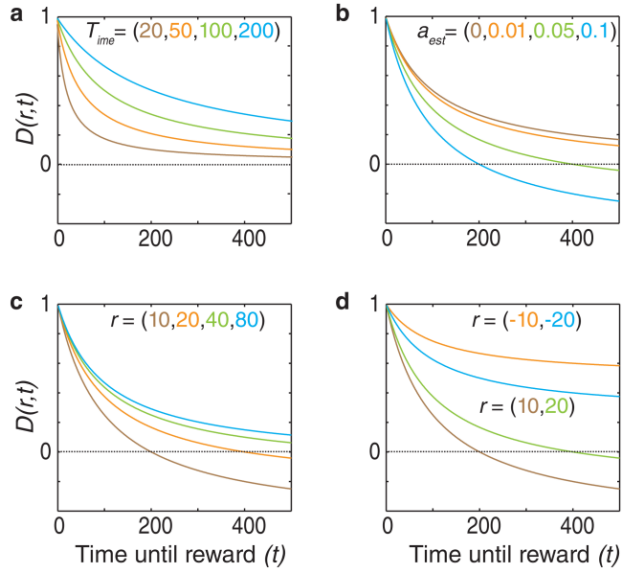

**Figure 3:** *The dependence of the discounting function on its parameters (Equation 3).* **a**) *Explicit temporal cost of waiting:* As the past integration interval ($T_{ime}$) increases, the discounting function becomes less steep, i.e. the subjective value for a given delayed reward becomes higher ($a_{est} = 0$ and $r = 20$). **b**) *Opportunity cost affects discounting:* As $a_{est}$ increases, the opportunity cost of pursuing a delayed reward increases and hence, the discounting function becomes steeper. The dotted line indicates a subjective value of zero, below which rewards are not pursued, as is the case when the delay is too high. ($r = 20$ and $T_{ime} = 100$). **c**) *"Magnitude Effect":* As the reward magnitude increases, the steepness of discounting decreases[3,4,10] ($T_{ime} = 100$ and $a_{est} = 0.05$). **d**) *"Sign Effect" and differential treatment of losses:* Gains (green and brown) are discounted steeper than losses (cyan and orange) of equal magnitudes[3,4] ($T_{ime} = 100$ and $a_{est} = 0.05$). Note that as the magnitude of loss decreases, so does the steepness of discounting (Supplementary Figure S3). In fact, for losses with magnitudes lower than $a_{est}T$, the discounting function will be greater than 1, leading to a differential treatment of losses[3,4] (see text, Figure S3).

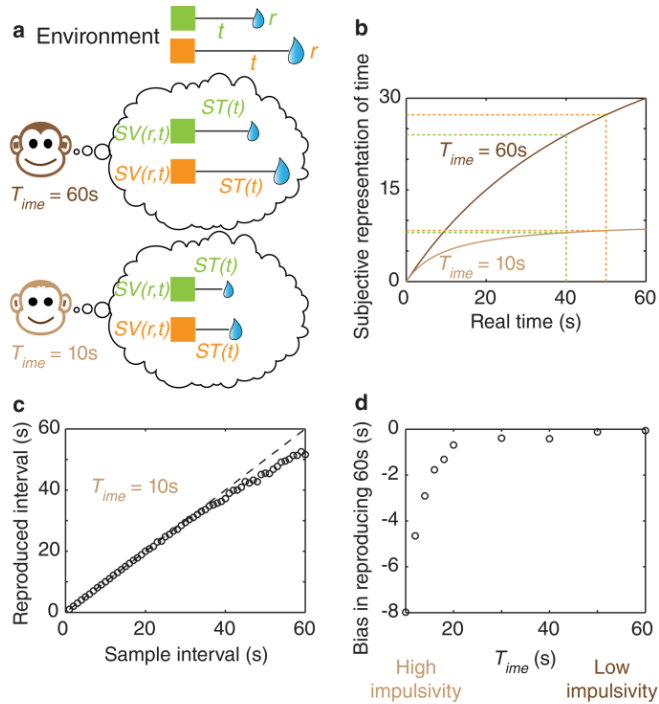

**Figure 4:** *Subjective time mapping and simulations of performance in a time reproduction task.* **a)** A schematic of the representation of the reward-environment by two animals with different values of $T_{ime}$. Lower values of $T_{ime}$ generate steeper discounting (higher impulsivity), and hence, smaller subjective values. **b)** *Subjective time mapping:* The subjective time mapping as expressed in *Equation 5* is plotted for the two animals in **a**. Subjective time representation saturates at $T_{ime}$ for longer intervals. This saturation effect is more pronounced in the case of higher impulsivity, thereby leading to a reduced ability to discriminate between intervals (here, 40 and 50s). **c)** *Bias in time reproduction:* A plot of reproduced median intervals for a case of high impulsivity in a simulated time reproduction task as generated by the simple accumulator model (see *Methods*; Figure S2) for sample intervals ranging between 1 and 60s. At longer intervals, there is an increasing underproduction. **d)** The bias in timing (difference between reproduced interval and sample interval) a 60s sample interval is shown for different values of $T_{ime}$, demonstrating that as impulsivity reduces, so does underproduction.

**SUPPLEMENTARY INFORMATION**

**Table of Contents**

# 1 RESULTS

## 1.1 TIMERR theory of decision making

Intertemporal choice behavior has been modeled using two dissimilar approaches. The first approach is to develop theories that explore ultimate(*1*) causes of behavior through general optimization criteria(*2–7*). In ecology, there are two dominant theories of intertemporal choice, Optimal Foraging Theory (OFT) and Ecological Rationality Theory (ERT). OFT posits that the choice behavior of animals results from a global maximization of a "fitness currency" representing long-term reward rates(*4*, *5*). Therefore, OFT predicts a temporal discounting function— the ratio of subjective value of a delayed reward to the subjective value of the reward when presented immediately—that decays hyperbolically with the delay. ERT states that it is sufficient to maximize reward rates only over the choice under consideration, (i.e. locally) to attain ecological success(*5–7*). ERT also predicts a hyperbolic discounting function. In economics, Discounted Utility Theory (DUT)(*2*, *3*) posits that animals maximize long-term exponentially-discounted utility so as to maintain temporal consistency of choice behavior(*2*, *3*).

The second approach, mainly undertaken by psychologists and behavioral analysts, is to understand the proximate(*1*) origins of choices by modeling behavior using empirical fits to data collected from standard laboratory tasks(*8*). An overwhelming number of these behavioral experiments, however, contradict the above theoretical models. Specifically, animals exhibit hyperbolic discounting functions, inconsistent with DUT(*3*, *5*, *8*, *9*), and violate the postulate of global reward rate maximization, inconsistent with OFT(*5*, *6*, *8*, *9*). Further, there are a wide variety of observations like 1) data being better-fit by a discounting function that is steeper than a pure-hyperbolic function(*3*, *8*), 2) the variability of discounting steepness within and across individuals(*3*, *10*, *11*), and many "anomalous" behaviors including 3) "Magnitude Effect" (*3*, *8*), 4) "Sign Effect"(*3*, *8*) and 5) differential treatment of punishments(*3*, *8*, *12*), that are not explained by ERT. It must also be noted that none of the above theories are capable of explaining how animals measure delays to rewards, nor do prior theories of time perception(*13*, *14*) attempt to explain intertemporal choice. Though psychology and behavioral sciences attempt to rationalize the above observations by constructing proximate models invoking phenomena like attention, memory and mood(*3*, *8*, *15*), ultimate causes are rarely proposed. As a consequence, these models of animal behavior are less parsimonious, and often ad-hoc.

In order to explain behavior, an ultimate theory must consider appropriate proximate constraints. The lack of appropriate constraints might explain the inability of the above theories to rationalize experimental data. By merely stating that animals maximize indefinitely-long-term reward rates or discounted utility, OFT and DUT expect animals to consider the effect of all possible reward-options in the past and the future when making the current choice(*4*, *8*) (Figure S1). However, such a solution would be biologically implausible for at least three reasons: 1) animals cannot know all the rewards obtainable in the future; 2) even if animals knew the disposition of all possible future rewards, the combinatorial explosion of such a calculation would present it with an untenable computation (e.g. in order to be optimal when performing even one hundred sequential binary choices, an animal will have to consider each of the $2^{100}$ combinations); 3) animals cannot persist for indefinitely long intervals without food in the hope of obtaining an unusually large reward in the distant future, even if the reward may provide the highest long-term reward rate (e.g. option between 11000 units of reward in 100 days vs. 10 units of reward in

0.1 day). On the other hand, ERT, although computationally-simple, expects an animal to ignore its past reward experience while making the current choice (Figure S1).

To contend with uncertainties regarding the future, an animal could estimate reward rates based on an expectation of the environment derived from its past experience. Specifically, instead of maximizing reward rates over an indefinite time into the future, an animal could choose rewards that increase its reward rate using its immediate past. In so doing, an animal can estimate the opportunity costs involved in pursuing a delayed reward. In a world that presents large fluctuations in reinforcement statistics over time, estimating reinforcement rate using the immediate past has an advantage over using longer-term estimations because the correlation between immediate past and the immediate future is likely high.

Hence, in order to maximize biologically-relevant estimates of reward rates, a computationally simplistic calculation is to maximize reward rates over a constrained temporal window. Specifically, this temporal window can span a past integration interval ($T_{ime}$) (over which past reward rate is estimated) and the expected delay to a future reward. It must be noted that such a simple algorithm also reduces the likelihood of waiting prohibitively long delays for large rewards (*Supplementary Information-2.2*).

Suppose that the animal is presented with $k$ options, each denoted by a label $i$. Picking the option $i$ will present the animal with a reward $r_i$ at a delay of $t_i$. Let's denote the $i^{th}$ option by the pair of variables ($r_i$, $t_i$). The animal has to make a choice among the $k$ options so as to maximize the global reward rate in the temporal window of consideration ($T_{ime}+t_i$).

If we denote the estimated average reward rate appropriate for the past integration window of $T_{ime}$ to be $a_{est}$, the global reward rate integrated over the duration of $T_{ime}+t_i$ for option i would be given by:

$$global\ reward\ rate = \frac{total\ reward\ earned}{total\ time\ spent} = \frac{a_{est}T_{ime}+r_i}{T_{ime}+t_i}$$

TIMERR theory (Figure 1 in main text) postulates that the animal will pick option $i_g$ in order to maximize the global reward rate, where

$$i_g = \max_{i} \frac{a_{est}+\frac{r_i}{T_{ime}}}{1+\frac{t_i}{T_{ime}}} \qquad (Equation\ S.1)$$

A feature of this decision rule is that delays following the reward are not considered. This is a consequence of the representation of the delayed reward: only delays up to the reward are stored in memory. However, post-reward delays are implicitly included in the estimation of past reward rate.

## 1.2  Calculation of subjective value and discounting function

The subjective value of a delayed reward option *(r,t)* is defined as the amount of immediate reward that is the subjective equivalent to the delayed option. According to the decision rule of

TIMERR theory, subjective equivalence is attained when two delayed rewards lead to equal reward rates. Hence, the subjective value of an option will obey the following equation:

$$\text{Reward rate for } (SV(r,t),0) = \text{Reward rate for } (r,t)$$

i.e.

$$\frac{a_{est} + \dfrac{SV(r,t)}{T_{ime}}}{1 + \dfrac{0}{T_{ime}}} = \frac{a_{est} + \dfrac{r}{T_{ime}}}{1 + \dfrac{t}{T_{ime}}}$$

where $SV(r,t)$ is the subjective value of reward $r$ delayed by time $t$. Simplifying, the expression for $SV(r,t)$ is given by

$$SV(r,t) = \frac{r - a_{est}t}{1 + \dfrac{t}{T_{ime}}} \qquad (Equation\ S.2)$$

The decision rule of TIMERR theory can also be expressed as picking the option with the highest subjective value as calculated above. The above expression for subjective value illustrates that the attribution of value can be re-conceptualized as the subtraction of the amount of reward the animal expects to lose by waiting the delay ($a_{est}t$) from the expected reward ($r$). This net gain is then divided by the cost of waiting the delay.

It should also be noted that the above decision rule can be applied even when deciding whether to pursue or forgo a single delayed reward. So long as the subjective value is negative (positive) for such a reward, it should be forgone (pursued).

Prior literature has modeled subjective value of a delayed reward as the product of a discounting function and the reward magnitude. Hence, the discounting function is the ratio of subjective value to the subjective value of the reward when presented immediately, i.e.

$$D(r,t) = \frac{SV(r,t)}{r} = \frac{1 - \dfrac{a_{est}}{r}t}{1 + \dfrac{t}{T_{ime}}} \qquad (Equation\ S.3)$$

where $D(r,t)$ is the discounting function for a reward $r$ delayed by a time $t$.

The above discounting function is hyperbolic in form with an additional subtractive term indicating the opportunity cost. It must be noted that the temporal cost imposed on waiting for a delayed reward depends intimately on the ratio of that delay to the past integration window ($t/T_{ime}$). Hence, if $T_{ime}$ is relatively large, the discounting is increasingly shallow and the animal would display high amounts of tolerance to delay in waiting for a delayed reward, whereas if $T_{ime}$ is relatively small, the discounting will be steeper and the animal would be less tolerant to delay in waiting for a delayed reward.

## 1.3 Extensions of TIMERR theory

### 1.3.1 Alternative version (store reward rates evaluated upon the receipt of reward in memory)

An alternative version of TIMERR theory could be appropriate for very simple life forms with limited computational resources that are capable of intertemporal decision making (e.g. insects). Rather than representing both the magnitude and delay to rewards separately and making decisions based on real time calculations, upon the receipt of reward, such animals could store subjective value directly in memory. In such a case, the reward rate at the time of reward receipt would be calculated over $T_{ime}+t$ and converted to subjective value. The decision between reward options is then simply described as picking the option with the highest stored subjective value. Mathematically, such a calculation is exactly equivalent to the calculation presented in *Supplementary Information-1.1*.

While the advantage of this model is that it is computationally less expensive, the disadvantages for the model are that 1) subjective values in memory are not generalizable, i.e. the subjective value in memory for an option will fundamentally depend on the reward environment in which it was presented; and 2) representations of the reward delays could be useful for anticipatory behaviors.

### 1.3.2 Evaluation of risk

Until now, we have assumed that a delayed reward *will* be available for consumption, provided the animal waits the delay, i.e. there are no explicit risks in obtaining the reward. In many instances in nature, however, such an assumption is not true. If the animal could build a model of the risks involved in obtaining a delayed reward, it could do better by including such a model in its decision making. Given information about a delayed reward $(r,t)$, if the animal could predict the expected reward available for consumption after having waited the delay ($ER(r,t)$), the subjective value of such a reward could be written as

$$SV(r,t) = \frac{ER(r,t) - a_{est} t}{1 + \frac{t}{T_{ime}}} \qquad (Equation\ S.4)$$

This is based on *Equation S.2*.

It is important to note that this equation can still be expressed in terms of subjective time as defined in the Main Text, viz.

$$SV(r,t) = \left( \frac{ER(r,t)}{t} - a_{est} \right) \frac{t}{1 + \frac{t}{T_{ime}}}$$

Generally speaking, building such risk models is difficult, especially since they are environment-specific. However, there could be statistical patterns in environments for which animals have acquired corresponding representations over evolution. Specifically, decay of rewards arising from factors like natural decay (rotting, for instance) or due to competition from other foragers could have statistical patterns. During the course of travel to a food source, competition poses the strongest cause for decay since natural decay typically happens over a longer time-scale, viz. days to months. In such an environment with competition from other foragers, a forager could estimate how much a reward will decay in the time it takes it to travel to the food source.

Suppose the forager sees a reward of magnitude $r$ at time $t = 0$, the moment of decision. The aim of the forager is to calculate how much value will be left by the time it reaches the food source, and to use this estimate in its current decision. Let us denote the time taken by the forager to travel to the food source by $t$.

We assume that the rate of decay of a reward in competition is proportional to a power of its magnitude, implying that larger rewards are more sought-after in competition and hence, would decay at a faster rate. We denote the survival time of a typical reward by $t_{sur}$ and consider that after time $t_{sur}$, the reward is entirely consumed. If, as stated above, one assumes that $t_{sur}$ is inversely related to a power $\alpha$ of the magnitude of a reward at any time ($r(t)$), we can write that

$$t_{sur} = \frac{1}{kr(t)^\alpha}$$ where $k$ is a constant of proportionality.

Hence, the rate of change of a value with initial magnitude $r$, will be

$$\frac{dr(t)}{dt} = -\frac{r(t)}{t_{sur}} = -\left(kr(t)^\alpha\right)r(t)$$

Solving this differential equation for $r(t)$,

$$r(t) = \frac{r}{\left(1+k\alpha r^\alpha t\right)^{1/\alpha}}$$

Here we set $r(0) = r$.

A forager could estimate the parameters $k$ and $\alpha$ based on the density of competition and other properties of the environment. In such a case, the subjective value of a delayed reward $(r,t)$ should be calculated as

$$SV(r,t) = \frac{\dfrac{r}{\left(1+k\alpha r^\alpha t\right)^{1/\alpha}} - a_{est}t}{1+\dfrac{t}{T_{ime}}} \qquad \text{(Equation S.5)}$$

The discounting function in this case is

$$D(r,t) = \frac{\dfrac{1}{\left(1+k\alpha r^\alpha t\right)^{1/\alpha}} - \dfrac{a_{est}t}{r}}{1+\dfrac{t}{T_{ime}}} \qquad \text{(Equation S.6)}$$

Such a discounting function can be thought of as a quasi-hyperbolic discounting function, and is a more general form than *Equation S.2* since *k = 0* returns *Equation S.2*.

### 1.3.3 Non-linearities in subjective value estimation

Animals do not perceive rewards linearly (e.g. 20 liters of juice is not 100 times more valuable than 200 mL). Non-linear reward perception may reflect the non-linear utility of rewards: too little is often insufficient while too much is unnecessary. Further, the value of a reward depends on the internal state of an animal (e.g. 200 mL of juice is more valuable to a thirsty animal than a satiated animal). We address such non-linearities as applied to TIMERR theory here.

If the non-linearities and state-dependence of magnitude perception can be expressed by a function $f(r,state)$, then this function can be incorporated into *Equation S.2* to give

$$SV(r,t) = \frac{f(r,state) - a_{est}t}{1 + \frac{t}{T_{ime}}} \qquad \text{(Equation S.7)}$$

The introduction of such state-dependence and non-linearities may account for the anomalous "preference for spread"(*3, 8*) and "preference for improving sequences"(*3, 8*) seen in human choice behavior.

### 1.3.4 Expected reward rate gain during the wait

We have not yet considered the possibility that animals could expect to receive additional rewards during the wait to delayed rewards, i.e. while animals expect to lose an average reward rate of $a_{est}$ during the wait, there could be a different reward rate that they might, nevertheless, expect to gain. If we denote that this additional expected reward rate is a fraction *f* of $a_{est}$, then we can state that the net expected loss of reward rate during the wait is $(1-f)a_{est}$. This factor can also be added to expressions of subjective value calculated above in *Equations S.2, S.5 and S.7*. Specifically, *Equation S.2* becomes

$$SV(r,t) = \frac{r - (1-f)a_{est}t}{1 + \frac{t}{T_{ime}}} \qquad \text{(Equation S.8)}$$

Such a factor is especially important in understanding prior human experiments. In abstract questions like "$100 now or $150 a month from now?", human subjects expect an additional reward rate during the month and are almost certainly not making decisions with the assumption that the only reward they can obtain during the month is $150.

### 1.3.5 State-dependence of discounting steepness

In the basic version of TIMERR theory, the time window over which the algorithm aims to maximize reward rates is the past integration interval ($T_{ime}$) plus the time to a delayed reward. However, non-linearities in the relationship between reward rates and fitness levels (as discussed in *Supplementary Information-2.2*) could lead to state-dependent consumption requirements. For example, in a state of extreme hunger, it might be appropriate for the decision rule to apply a very short time scale of discounting so as to avoid dangerously long delays to food. However, integrating past reward rates over such extremely short timescales could compromise the reliability of the estimated reward rate. Hence, as a more general version of TIMERR theory, the window over which reward rate is maximized could incorporate a scaled down value of the interval over which past reward rate is estimated, with the scaling factor governed by consumption requirements. If such a scaling factor is represented by *s(state)*, *Equation S.2* would become

$$SV(r,t) = \frac{r - a_{est}t}{1 + \dfrac{t}{T_{ime}s(state)}} \quad (Equation\ S.9)$$

*1.3.6   Generalized expression for subjective value*

Combining *Equations S.5, S.7, S.8 and S.9*, we can write a more general expression for the subjective value of a delayed reward, including a model of risk along with additional reward rates, state dependences and non-linearities in the perception of reward magnitude

$$SV(r,t) = \frac{\dfrac{f(r, state)}{\left(1 + k\alpha f(r, state)^{\alpha} t\right)^{1/\alpha}} - (1-f)a_{est}t}{1 + \dfrac{t}{T_{ime}s(state)}} \quad (Equation\ S.10)$$

*Equation S.10* is a more complete expression for the subjective value of delayed rewards. Such an expression could capture almost the entirety of experimental results, considering its inherent flexibility. However, it should be noted that even with as simple an expression as *Equation S.2*, many observed experimental results can be explained.

**1.4    Temporal bisection experiments**

In temporal bisection experiments, a subject is first required to learn two template intervals, one being a short interval ($t_s$) and the other, a long interval ($t_l$). Once these reference intervals are learned, the subject is presented with sample intervals of durations between the short and the long templates. The subject is asked to classify the sample duration as closer to the short interval or to the long interval. The sample interval at which subjects show maximum uncertainty in classification as short or long is called the point of subjective equality, or, the "bisection point". The bisection point is of considerable theoretical interest for the following reasons. If subjects perceived time linearly with constant errors, the bisection point would be the arithmetic mean of the short and long intervals. On the other hand, if subjects perceived time in a scalar or logarithmic fashion, it has been proposed that the bisection point would be at the geometric mean(*16*). However, experiments studying temporal bisection have produced ambiguous results. Specifically, the bisection point has been shown to vary between the geometric mean and the arithmetic mean and has sometimes even been shown to be below the geometric mean, closer to the harmonic mean(*17*).

The bisection point as calculated by TIMERR theory is derived below. The calculation involves transforming both the short and long intervals into subjective time representations and expressing the bisection point in subjective time (subjective bisection point) as the mean of these two subjective representations. The bisection point expressed in real time is then calculated as the inverse of the subjective bisection point.

$$ST(t_s) = \frac{t_s}{1+\frac{t_s}{T_{ime}}} ; ST(t_l) = \frac{t_l}{1+\frac{t_l}{T_{ime}}}$$

Therefore, the bisection point in subjective time is given by

$$Subjective\ bisection\ point\ (SBP) = \frac{ST(t_s) + ST(t_l)}{2} = \frac{\frac{t_s}{1+\frac{t_s}{T_{ime}}} + \frac{t_l}{1+\frac{t_l}{T_{ime}}}}{2}$$

The value of the bisection point expressed in real time is given by the inverse of the subjective bisection point, viz.

$$Bisection\ point\ in\ real\ time = \frac{SBP}{1-\frac{SBP}{T_{ime}}} = \frac{T_{ime}\left(\frac{t_s+t_l}{2}\right) + t_s t_l}{T_{ime} + \left(\frac{t_s+t_l}{2}\right)} \quad (Equation\ S.11)$$

$$Bisection\ point = \frac{T_{ime}\left(\frac{t_s+t_l}{2}\right) + t_s t_l}{T_{ime} + \left(\frac{t_s+t_l}{2}\right)}$$

From the above expression, it can be seen that the bisection point can theoretically vary between the harmonic mean and the arithmetic mean as $T_{ime}$ varies between zero and infinity respectively.

Hence, TIMERR theory predicts that when comparing bisection points across individuals, individuals with larger values of $T_{ime}$ will show bisection points closer to the arithmetic mean whereas individuals with smaller values of $T_{ime}$ will show lower bisection points, closer to the geometric mean. If $T_{ime}$ was smaller still, the bisection point would be lower than the geometric mean, approaching the harmonic mean. This is in accordance with the experimental evidence referenced above. Further, we also predict that the steeper the discounting function, the lower the bisection point, as has been experimentally confirmed(*18*).

### 1.5 Errors in timing

The major predictions of TIMERR theory in relation to time perception are that 1) the accuracy in time representations decreases for longer intervals; 2) such a decrease is larger in magnitude when impulsivity is high and that 3) impulsivity leads to an underproduction of time intervals.

However, it must be noted that the precise quantitative details of the above predictions (viz. dependence of timing accuracy on interval duration; dependence of the magnitude of underproduction on degree of impulsivity) will depend fundamentally on the proximate implementation of the expression for subjective time as shown in *Equation 5* of the main text.

Assuming a simple accumulator model for the proximate implementation of *Equation 5*, the error in representation of an interval can be shown to grow quadratically with the duration of the interval. This is because if one assumes that the representation of subjective time, *ST(t)*, has a constant infinitesimal noise of *dST(t)* associated with it, the noise in representation of a true interval *t*, denoted as *dt* will obey

$$\frac{dST(t)}{dt} = \frac{d}{dt}\left(\frac{t}{\left(1+\frac{t}{T_{ime}}\right)}\right) = \frac{1}{\left(1+\frac{t}{T_{ime}}\right)^2}$$

If one assumes that the neural noise in representing *ST(t)* is a constant *c*, independent of *t*, then the corresponding error in real time is

$$dt = c\left(1+\frac{t}{T_{ime}}\right)^2$$

The coefficient of variation (error/central tendency) expected from such a model is then

$$Cv \approx \frac{c\left(1+\frac{t}{T_{ime}}\right)^2}{t} = c\left(\frac{1}{t}+\frac{2}{T_{ime}}+\frac{t}{T_{ime}^2}\right) \qquad (Equation\ S.12)$$

This is a U-shaped curve since, as time increases, it will first decrease, then appear constant over a limited range, and then increase linearly as time further increases.

The assumption of just one accumulator and a decoder may be overly simplistic for an animal brain. If one assumes an ensemble of accumulators, each tuned to a slightly different integration constant, the decoding and subsequent dependence of errors on the duration represented could deviate significantly from the simple expression shown above. As a consequence, the linear range over which scalar timing is observed would be larger. Also, even with a single accumulator, a more realistic model representing neural processing would incorporate a mean-reverting process rather than a simple Wiener process considered here, leading to further deviations from the above expression. Such involved calculations are beyond the scope of this work. Nevertheless, independent of how *Equation 5* is implemented in a proximate model, the three enumerated qualitative predictions regarding errors in timing will hold.

## 2 DISCUSSION

### 2.1 Implications for intertemporal choice

*2.1.1 Consequences of the discounting function*

We rewrite *Equation S.3* (*Equation 3* in Main text) below followed by its implications for intertemporal choice in environments with positive and negative past reward rate estimates.

$$D(r,t) = \frac{SV(r,t)}{r} = \frac{1 - \frac{a_{est}}{r}t}{1 + \frac{t}{T_{ime}}}$$

In an environment with positive $a_{est}$, the following predictions can be made

1. *"Magnitude Effect" for gains*: as noted in the Main Text, as $r$ increases, the numerator increases in value, effectively making the discounting less steep (Figure 3c). This effect has been experimentally observed and has been referred to as the "magnitude" effect(*3, 8*). TIMERR theory makes a further prediction, however, that the size of the "magnitude" effect will depend on the size of $a_{est}$ and $t$. Specifically, as $a_{est}$ and $t$ increase, so does the size of the effect.
2. *"Magnitude Effect" for losses/punishments*: if $r$ is negative (i.e. loss/punishment), the discounting function will become more steep as the magnitude of $r$ increases (Figure 3d, Figure S3). Hence, in a rewarding environment ($a_{est} > 0$), the "magnitude" effect for punishments is in the opposite direction as the "magnitude" effect for gains.
3. *"Sign Effect"*: gains are discounted more steeply than punishments of equal magnitudes. A further prediction is that this effect will be larger for smaller reward magnitudes. This prediction has been proven experimentally(*3, 12*).
4. *Differential treatment of losses/punishments*: As the "magnitude" of the punishment decreases below $a_{est}T_{ime}$ ($r > -a_{est}T_{ime}$), the discounting function becomes a monotonically increasing function of delay. This means that the punishment would be preferred immediately when the magnitude of punishment is below this value. Above this value, a delayed punishment would be preferred to an immediate punishment. This prediction has experimental support(*3, 8*).
5. A reward of $r$ delayed beyond $t = r/a_{est}$ will lead to a negative subjective value. Hence, given an option between pursuing or forgoing this reward, the animal would only pursue (forgo) the reward at shorter (longer) delays.

When understanding the reversal of the "Magnitude Effect" for losses, it is important to keep in mind that as $|r| \to \infty$, both losses and gains approach the same asymptote.

$$D(r,t;|r| \to \infty) = \frac{1}{1 + \frac{t}{T_{ime}}}$$

Hence, as the magnitude of a loss increases, the size of the "Magnitude Effect" becomes lower and harder to detect (Figure S3).

In an environment with negative $a_{est}$ (i.e. net punishing environment), all the predictions listed above would reverse trends. Specifically,

6. *"Magnitude Effect" for gains*: as $r$ increases, the discounting becomes steeper
7. *"Magnitude Effect" for losses*: as the magnitude of a punishment increases, the discounting function becomes less steep.
8. *"Sign Effect"*: Punishments are discounted more steeply than gains of equal magnitudes.
9. *Differential treatment of gains*: as the magnitude of the gain decreases below $a_{est}T_{ime}$ (r < $-a_{est}T_{ime}$), it would be preferred at a delay. Beyond this magnitude, the gain would be preferred immediately.
10. A punishment of magnitude $r$ will be treated with positive subjective value if it is delayed beyond $t = r/a_{est}$.

*2.1.2 Animals do not maximize long-term reward rates*

In typical animal intertemporal choice experiments, in order to ensure that different reward options do not lead to a marked difference in overall experiment duration, a post-reward delay is introduced for all options such that the net duration of each trial is constant. In such experiments, a global-reward-rate-maximizing agent should always choose the larger reward, irrespective of the cue-reward delay, since the net time spent per trial in collecting any reward equals the constant trial duration. However, a preponderance of experimental evidence shows that animals deviate from such ideal behavior of maximizing reward rates over the entire session(*5, 6, 8*). Such experimental results are typically interpreted to signify that animals do not, in fact, act as reward-rate-maximizing agents(*5, 6, 8*). TIMERR theory proposes that even though animals are maximizing reward rates, albeit under constraints of experience, post-reward delays are not incorporated into their decision process due to limitations of associative learning(*19*). As a consequence, animal choice behavior in such laboratory tasks would not maximize global reward rates.

TIMERR theory, however, allows for the possibility that in a variant of standard laboratory tasks that makes a post-reward delay immediately precede another reward included in the choice behavior would result in animals not ignoring post-reward delays. Prior experiments evince this possibility(*6*). Specifically, post-reward delays are included in the decision process by birds performing a patch leave-stay task that is economically equivalent to standard laboratory tasks on intertemporal choice(*6*). Since the exclusion of post-reward delays in decisions is borne out of limitations of associative learning, TIMERR theory also allows for the inclusion of these delays in tasks where they can be learned. Presumably, an explicit cue indicating the end of post-reward delays could foster a representation and inclusion of these delays in decisions. It has been shown in a recent experiment that monkeys include post-reward delays in their decisions when they are explicitly cued(*9*).

## 2.2 Effects of plasticity in the past integration interval ($T_{ime}$)

The most important implication of the TIMERR theory is that the steepness of discounting of future rewards will depend directly on the past integration interval, i.e. the longer you integrate over the past, the more tolerant you will be to delays, and vice-versa. In the above sections, the past integration interval ($T_{ime}$) was treated as a constant. However, the purpose of the past integration interval is to reliably estimate the baseline reward rate expected through the delay until a future reward. Further, since $T_{ime}$ determines the temporal discounting steepness, it will also affect the rate at which animals obtain rewards in a given environment. Hence, depending on the reinforcement statistics of the environment, it would be appropriate for animals to adaptively integrate reward history over different temporal windows so as to maximize rates of reward.

In this section, we qualitatively address the problem of optimizing $T_{ime}$. We consider that an optimal $T_{ime}$ would satisfy four criteria: 1) obtain rewards at magnitudes and intervals that maximize the fitness of an animal, which is accomplished partially through 2) reliable estimation of past reward rates leading to 3) appropriate estimations of opportunity cost for typical delays faced by the animal with 4) minimal computational/memory costs.

Before considering the general optimization problem for $T_{ime}$, it is useful to consider an illustrative example. This example ignores the last three criteria listed above and only considers the impact of $T_{ime}$ on the fitness of an animal. Consider a hypothetical animal that typically obtains rewards at a rate of 1 unit per hour. Suppose such an animal is presented with a choice between a) 2 units of reward available after an hour, and b) 20 units of reward available after 15 hours. The subjective values of options 'a' and 'b' are calculated below for four different values of $T_{ime}$, as per *Equation S.2*.

|  | Subjective value of 'a' | Subjective value of 'b' | Chosen option |
|---|---|---|---|
| $T_{ime} = \infty$ hours | 1 | 5 | b |
| $T_{ime} = 10$ hours | 0.91 | 2 | b |
| $T_{ime} = 2.5$ hours | 0.71 | 0.71 | Both equal |
| $T_{ime} = 1 hour$ | 0.50 | 0.31 | a |

As is apparent, larger $T_{ime}$ biases the choice towards option 'b'. This is appropriate in order to maximize long term reward rate since the long term reward rate is higher for option 'b', as shown below.

Reward rate having chosen option 'b' = 20 units in 15 hours = 20/15 units/hour.

Reward rate having chosen option 'a' = 2 units in one hour + 14 units in the remaining 14 hours = 16/15 units/hour.

However, if we presume that this animal evolved so as to require a minimum reward of 2 units within every 10 hours in order to function in good health, choosing option 'b' would be inappropriate. Hence, it is clear that for this hypothetical animal, $T_{ime}$ should be much lower than 10 hours. In summary, so as to meet consumption requirements, it is inappropriate to integrate past reward rate history over very long times even if the animal has infinite computational/memory resources. Keeping in mind the above example and the four criteria listed for an optimal $T_{ime}$, we enumerate the following disadvantages for setting inappropriately large or inappropriately small $T_{ime}$.

Integrating over inappropriately large $T_{ime}$ has at least four disadvantages to the animal: 1) a very long $T_{ime}$ is inappropriate given consumption requirements of an animal, as illustrated above; 2) the computational/memory costs involved in this integration are high; 3) integrating over large time scales in a dynamically changing environment could make the estimate of past reward rate inappropriate for the delay to reward (e.g. integrating over the winter and spring seasons as an estimate of baseline reward rate expected over a delay of an hour in the summer might prove very costly for foragers); 4) the longer the $T_{ime}$, the harder it is to update $a_{est}$ in a dynamic environment.

Integrating over inappropriately small $T_{ime}$, on the other hand, presents the following disadvantages: 1) estimate of baseline reward rate would be unreliable since integration must be carried out over a long enough time-scale so as to appreciate the stationary variability in an environment; 2) estimate of baseline reward rate might be highly inappropriate for the future delay (e.g. integrating over the past one minute might be very inappropriate when the delay to a future reward is a day); 3) the animal would more greatly deviate from global optimality (as is clear from *Equation S.2*).

In light of the above discussion, we argue that the following relationships should hold for $T_{ime}$. In each of these relationships, all factors other than the one considered are assumed constant.

R1. *Time-dependent changes in environmental reinforcement statistics:* if an environment is unstable, i.e. the reinforcement statistics of the environment are time- dependent, we predict that $T_{ime}$ would be lower than the timescale of the dynamics of changes in environmental statistics.

R2. *Variability of estimated reward rate:* if an environment is stable and has very low variability in the estimated reward rate it provides to an animal, integrating over a long $T_{ime}$ would not provide a more accurate estimate of past reward rate than integrating over a short $T_{ime}$. Hence, in order to be better at adapting to potential changes in the environment and minimize computational/ memory costs, we predict that in a stable environment, $T_{ime}$ will reduce (increase) as the variability in the estimated reward rate reduces (increases).

R3. *Mean of estimated reward rate:* in a stable environment with higher average reward rates, the benefit of integrating over a long $T_{ime}$ will be smaller when weighed against the computational/memory cost involved. As an extreme example, when the reward rate is infinity, the benefit of integrating over long windows is infinitesimal. This is because the benefit of integrating over a longer $T_{ime}$ can be thought of as the net gain in average reward rate over that achieved when decisions are made with the lowest possible $T_{ime}$. If the increase in average reward rate is solely due to an increase in the mean (constant standard deviation) of reward magnitudes, the proportional benefit of integrating over a large $T_{ime}$ reduces. If the increase in average reward rate is solely due to an increase in frequency of rewards, the integration can be carried out over a lower time to maintain the estimation accuracy. Hence, we predict that, in general, as average reward rates increase (decrease), $T_{ime}$ will decrease (increase).

R4. *Average delays to rewards:* as the average delay between the moment of decision and receipt of rewards increases (decreases), $T_{ime}$ should increase (decrease) correspondingly. This is because reward history calculated over a low $T_{ime}$ might be inappropriate as an estimate of baseline reward rate for the delays until future reward.

In human experiments, it is common to give abstract questionnaires to study preference (e.g. "which do you prefer: $100 now or $150 a month from now?"). In such tasks, setting $T_{ime}$ to be of the order of seconds or minutes might be very inappropriate to calculate a baseline expected reward rate over the month to a reward (*R4* above). Hence, we predict that $T_{ime}$ might increase so as to match the abstract delays to allow humans to discount less steeply as these delays increase.

Similarly, when the choice involves delays of the order of seconds, integrating over hours might not be appropriate and therefore, the discounting steepness would be predicted to be higher in such experiments. Thus, in prior experimental results(*3, 10–12*), $T_{ime}$ might have changed to reflect the delays queried.

## 2.3 Calculation of the estimate of past reward rate ($a_{est}$)

It must be noted that even though the calculation of $a_{est}$ is performed over a time-scale of $T_{ime}$, yet unspecified is the particular form of memory for past reward events. The simplest form of a memory function is one in which rewards that were received within a past duration of $T_{ime}$ are recollected perfectly while any reward that was received beyond this duration is completely forgotten. A more realistic memory function will be such that a reward that was received will be remembered accurately with a probability depending on the time in the past at which it was received, with the dependence being a continuous and monotonically decreasing function. For such a function, $T_{ime}$ will be defined as twice the average recollected duration over the probability distribution of recollection. The factor of two is to ensure that in the simplest memory model presented above, the longest duration at which rewards are recollected (twice the average duration) is $T_{ime}$.

If we define local updating as updating $a_{est}$ based solely on the memory of the last reward (both magnitude and time elapsed since its receipt), the constraint of local updating when placed on such a general memory function necessitates it to be exponential in time. In this case, $a_{est}$ is updated as:

$$a_{est} \to a_{est} + \frac{2r}{T_{ime}}; \text{ upon receipt of reward}$$

$$a_{est} \to a_{est} \exp(-\frac{2t_{lastreward}}{T_{ime}}); \text{ otherwise}$$

where $t_{lastreward}$ is the time elapsed since the receipt of the last reward.

## 2.4 Predictions of TIMERR theory supported experimentally

1. The discounting function will be hyperbolic in form(*3, 8*).
2. The discounting steepness could be labile within and across individuals(*3, 10–12, 15*).
3. Temporal discounting could be steeper when average delays to expected rewards are lower(*3, 10, 11*).
4. "Magnitude Effect": as reward magnitudes increase in a net positive environment, the discounting function becomes less steep(*3, 8*).
5. "Sign Effect": rewards are discounted steeper than punishments of equal magnitudes in net positive environments(*3, 8*).
6. The "Sign Effect" will be larger for smaller magnitudes(*3, 12*).
7. "Magnitude Effect" for losses: as the magnitudes of losses increase, the discounting becomes steeper. This is in the reverse direction as the effect for gains(*20*). Such an effect is more pronounced for lower magnitudes(*20*).
8. Punishments are treated differently depending upon their magnitudes. Higher magnitude punishments are preferred at a delay, while lower magnitude punishments are preferred immediately(*3, 8, 12*).

9. "Delay-Speedup" asymmetry: Delaying a reward that you have already obtained is more punishing than speeding up the delivery of the same reward from that delay is rewarding. This is because a received reward will be included in the current estimate of past reward rate ($a_{est}$) and hence, will be included in the opportunity cost(*3*, *8*).
10. Time perception and temporal discounting are correlated(*21*).
11. Timing errors increase with the duration of intervals(*13*, *14*, *22*, *23*).
12. Timing errors increase in such a way that the coefficient of variation follows a U-shaped curve(*14*, *24*).
13. Impulsivity (as characterized by abnormally steep temporal discounting) leads to abnormally large timing errors(*21*, *25*).
14. Impulsivity leads to underproduction of time intervals, with the magnitude of underproduction increasing with the duration of the interval(*21*).
15. The bisection point in temporal bisection experiments will be between the harmonic and arithmetic means of the reference durations(*16–18*).
16. The bisection point need not be constant within and across individuals(*18*).
17. The bisection point will be lower for individuals with steeper discounting(*18*).
18. The choice behavior for impulsive individuals will be more inconsistent than for normal individuals(*26*). This is because their past reward rate estimates will show larger fluctuations due to a lower past integration interval.
19. Post-reward delays will not be included in the intertemporal decisions of animals during typical laboratory tasks(*5*, *6*, *8*, *9*). Variants of typical laboratory tasks may, however, lead to the inclusion of post-reward delays in decisions(*5*, *6*, *8*, *9*).

**Supplementary References**

**Supplementary Figures:**

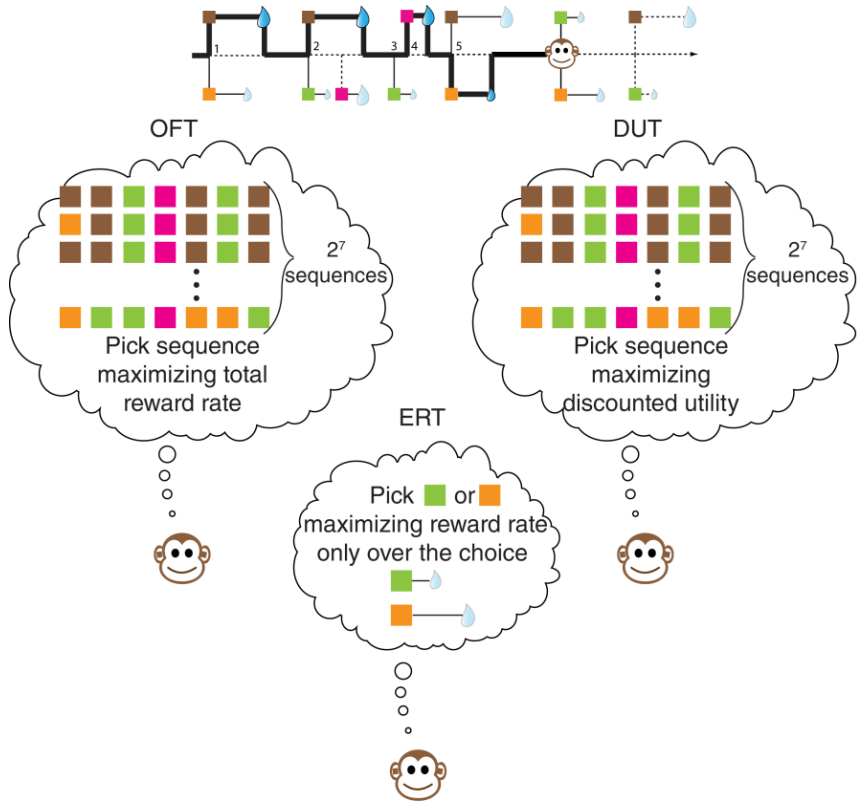

**Figure S1:** *A schematic illustration of the postulates of Optimal Foraging Theory, Discounted Utility Theory and Ecological Rationality Theory*: For the choice as shown (reproduced from Figure 1), both OFT and DUT states that the fitness variable of interest (reward rate for OFT and discounted utility for DUT) is maximized over the entire sequence of choices. Were the animal to know the disposition of every reward-option available in the future, it has to pick the sequence of options that will maximize the fitness variable of interest over all choices. Hence, in order to be optimal, the animal is forced to consider every possible sequence of reward-options including the future options (128 sequences for 7 total binary choices) so as to pick the optimal sequence. These algorithms assume that the animal can predict or otherwise know the future, at the moment of decision. ERT, on the other hand, states that the animal maximizes reward rate only over the current choice, thus ignoring the effects of prior reward experience. Further, a decision to forgo all options is not possible under ERT.

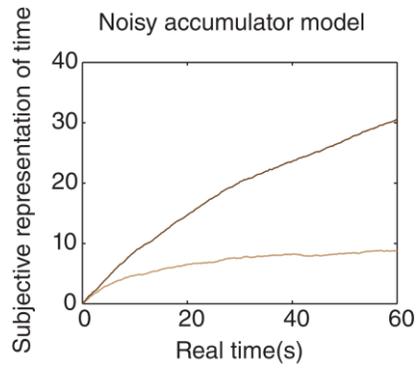

**Figure S2:** *Accumulation of the subjective representation of time using a noisy accumulator model*(see *Methods*): The subjective representation of time, as plotted in Figure 4B, is simulated using a noisy accumulator model with a Weiner process of 10% noise. The accumulated value is stored at the interval being timed (here 60s), stored in memory, and used as a threshold for later time reproduction. The reproduced interval (as in Figure 4C,D) ends at the moment of first threshold-crossing.

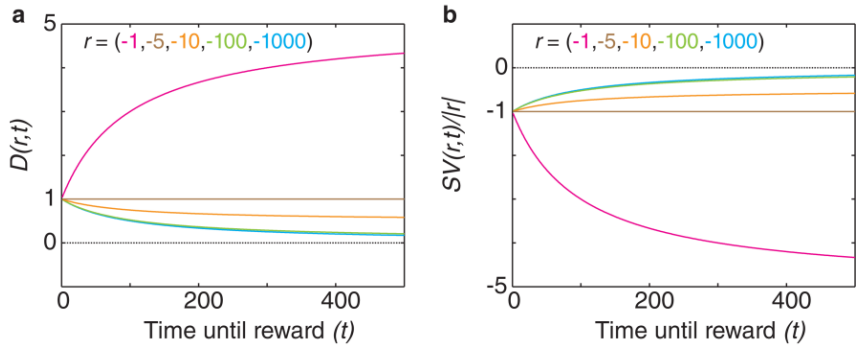

**Figure S3:** *"Magnitude Effect" and Differential treatment of losses in a net positive environment*: **a**) The discounting function plotted for losses of various magnitudes (as shown in Figure 3D; $a_{est} = 0.05$ and $T_{ime} = 100$). As the magnitude of a loss increases, the discounting function becomes steeper. However, the slope of the discounting steepness with respect to the magnitude is minimal for large magnitudes (100 and 1000; see *Supplementary Information-2.1.1*). At magnitudes below $a_{est}T_{ime}$, the discounting function becomes an increasing function of delay. **b**) Plot of the signed discounting function for the magnitudes as shown in **a**, showing that for magnitudes lower than $a_{est}T_{ime}$, a loss becomes even more of a loss when delayed. Hence, at low magnitudes ($<a_{est}T_{ime}$), losses are preferred immediately. No curve crosses the dotted line at zero, showing that at all delays, losses remain punishing.